\documentclass[final, a4paper, 10pt]{article}
\usepackage{natbib}
\usepackage[]{mcode}
\usepackage{amsmath}
\usepackage{amsthm}
\usepackage{amsfonts}
\usepackage{latexsym}
\usepackage{graphicx}
\usepackage{epstopdf}
\usepackage{color}
\DeclareMathAlphabet{\mathpzc}{OT1}{pzc}{m}{it}
\renewcommand{\Re}{\mathbb{R}}

\author{%
\\
\small
}
\date{ }
\title{Basic univariate and bivariate statistics for symbolic data: a critical review}
\begin{document}
\maketitle

\begin{abstract}
This few lines shows the main problem with the state-of-the-art of basic statistics for histogram and interval data.
\end{abstract}

\section{Numerical symbolic modal data}\label{sec:data}

[Histogram valued description] We assume that
$S(i)=[\underline{y}_i;\overline{y}_i]$ (the support is bounded in $\Re$).
The support is partitioned into a set of $n_i$ intervals $S(i)  = \left\{ {I_{1i}
,\ldots,I_{n_ii} } \right\}$, where  $ I_{hi}  = \left[
{\underline{y}_{hi} ,\overline{y}_{hi} } \right) $ and $h=1,\ldots,n_i$, i.e.
$$
\begin{array}{l}
i. \hspace{10pt}  I_{hi}  \cap I_{mi}  = \emptyset ;\hspace{5pt}h \ne m \hspace{5pt}; \\
ii. \hspace{10pt} \bigcup\limits_{l = 1,...,n_i} {I_{hi} }  = S(i) \\
 \end{array}
$$
 For histograms it is supposed that each interval is uniformly dense.
It is possible to define the modal description of $i$ as follows:
$$
Y(i) = \{ (I_{hi} ,\pi _{hi})\;|\;\forall I_{hi} \in S(i) ;\;\pi
_{hi}=\Phi_i(\underline{y}_{hi}\leq y \leq \overline{y}_{hi})  = \int\limits_{I_{hi}}  {\phi _{i} (z)dz}  \ge 0\}
$$
 where $\int\limits_{S(i)} {\phi _i (z)dz} =1$. In this case, the modal-numeric description is:
     $$
     Y(i)=\{(I_{1i},\pi_{1i}),\ldots,(I_{n_i i},\pi_{n_i i})\}.
     $$
 With $Y(i)$ it is possible to associated a distribution function $\Phi_i(y)$ as follows:
 $$\Phi_i(y)=\sum_{h<\ell}\pi_{hi}+\pi_{\ell i}\cdot \frac{y-\underline{y}_{\ell i}}{\overline{y}_{\ell i}-\underline{y}_{\ell i}}\;where\;(\ell\in{1,\ldots,n_i}:\overline{y}_{\ell i}\leq y\leq \underline{y}_{\ell i}).$$
 According to \cite{IrLeVe2006}, the corresponding quantile function (the inverse of $\Phi_i(y)$) is defined as :
 \begin{equation}\label{quantiHIST}
  \Phi_i^{-1}(t)=\underline{y}_{\ell i}+\frac{t-\Phi_i(\underline{y}_{\ell i})}{\pi_{\ell i}}\cdot(\overline{y}_{\ell i}-\underline{y}_{\ell i}),
 \end{equation}
  where $\left(\ell\in{1,\ldots,n_i}:\Phi_i(\underline{y}_{\ell i})\leq t\leq\Phi_i\left(\overline{y}_{\ell i}\right)\right).$

 \paragraph{Example} The pulse rate of the $i$ patient in a day is described through a histogram with support $S(i)=[80,120]$. The empirical frequency distribution of the observed pulse rate is described as a histogram as follows:
       $$Y(i)=\{([80,90),0.1), ([90,100), 0.3), ([100,110),0.4),([110,120],0.2)\}.
       $$
       It follows that $\Phi_i(y)$ is:
       $$\begin{array}{l|ccccc}
           I_{hi}\in S(i) & [80,90) & [90,100) & [100,110) & [110,120] \\
           \hline
           [\Phi_i(\underline{y}_{hi});\Phi_i(\overline{y}_{hi})] & [0, 0.1) & [0.1, 0.4) & [0,4, 0.8) &[0.8,  1]
         \end{array}. $$
         In this case, if we want to compute the quantile at level $t=0.5$ (i.e.,the median of the distribution) according to Eq. (\ref{quantiHIST}), we obtain that
     $$\Phi_i^{-1}(0.5)=100+\frac{0.5-0.4}{0.4}\cdot(110-100)=102.5\; .$$


\section{Basic univariate statistics for numerical symbolic data}\label{sec:Univariate_stats}
The first to propose a set of univariate and bivariate statistics for \emph{symbolic data} was \citet{BerGoup2000}, and subsequently \citet{BilDid2006} improved them. The \citet{BerGoup2000} approach  relies on the so-called \emph{two level paradigm} presented in SDA in \cite{BoDi2000}: the set-valued description of a statistical unit of a \emph{higher} order is the generalization of the values observed for a class of the \emph{lower} order units. For example, the income distribution of a \emph{nation} (the higher order unit) is the empirical distribution of the incomes of each citizen (the lower order units) of that nation. Naturally, other generalization of grouping criteria can be taken into consideration.\\
The generalization process from lower to higher order units considered by \citet{BerGoup2000} and by \citet{BilDid2006} implies the following assumptions: given two symbolic data $y(1)$ and $y(2)$ described by the frequency distributions $f_1(y)$ and $f_2(y)$, a lower order unit can be described by a single value $y_0$  that has a probability of occurring equal to $\frac{f_1(y_0)+f_2(y_0)}{2}$. The univariate statistics proposed by \citet{BerGoup2000} and by \citet{BilDid2006} for a symbolic variable (namely, a variable describing higher order units, or a class of units) correspond to those of the classic variable used for describing the (unknown) lower order units. Thus, given a set $E$ of $n$ higher order units described by the numerical symbolic variable $Y$, the mean, the variance and the standard deviation proposed by \citet{BerGoup2000} and extended by \citet{BilDid2006} correspond to those of a finite mixture of $n$ density (or frequency) functions with mixing weights equal to $\frac{1}{n}$. Given $n$ density functions denoted with $\phi_i(y)$ with the respective means $\mu_i=E(Y_i)$ and variance $\sigma^2_i=E[(Y_i-\mu_i)^2]$, and given the finite mixture density $\phi(y)$ as follows:
\begin{equation}\label{mixture}
  \phi(y)=\sum\limits_{i=1}^{n}{\frac{1}{n}\phi_i(y)}={\frac{1}{n}\sum\limits_{i=1}^{n}\phi_i(y)},
\end{equation}
\citet{Uhwir2006} shows that the mean $\mu=E(Y)$ and the variance $\sigma^2=E[(Y-\mu)^2]$ of $\phi(y)$ are the following:
\begin{eqnarray}
  \mu=E(Y)&=&\frac{1}{n}\sum\limits_{i=1}^{n}\mu_i; \label{mu_mix} \\
  \sigma^2=E[(Y-\mu)^2]&=&\frac{1}{n}\sum\limits_{i=1}^{n}{\left(\mu_i^2+\sigma_i^2\right)}-\mu^2. \label{s2_mix}
\end{eqnarray}
The \citet{Uhwir2006} mean and variance lead to the \citet{BilDid2006} mean and variance for an interval or a histogram symbolic variable. For the sake of simplicity, we show that this is true for interval-valued data. Indeed, the histogram-valued data are treated as a particular case of weighted interval descriptions. Let $Y$ be an interval-valued variable, thus, the generic symbolic data is $y(i)=[a_i;b_i]$ with $a_i \leq b_i$ belonging to $\Re$. According to \cite{BerGoup2000}, $y(i)$ is considered as a uniform distribution in $[a_i;b_i]$, with mean equal to  $\mu_i=\frac{a_i+b_i}{2}$ and variance equal to $\sigma^2_i=\frac{\left(b_i-a_i\right)^2}{12}$. Given a set of $n$ units described by an interval-valued variable, the {\em symbolic sample mean} $\bar{Y}$ \cite[eq. (3.22)]{BilDid2006} is:
\begin{equation}\label{mu_Bil}
\bar{Y}=\frac{1}{2\cdot n}\sum\limits_{i=1}^{n}{\left(b_i+a_i\right)}.
\end{equation}
It is straightforward to show its equivalence with $\mu$ in eq.(\ref{mu_mix}), indeed:
$$
\bar{Y}=\frac{1}{n}\sum\limits_{i=1}^{n}\frac{\left(b_i+a_i\right)}{2}=\frac{1}{ n}\sum\limits_{i=1}^{n}\mu_i=\mu .
$$
In \citep[eq. (3.22)]{BilDid2006} is also proposed the {\em symbolic sample variance} as follows:
\begin{equation}\label{var_Bi}
  S^2=\underbrace{\frac{1}{3\cdot n}\sum\limits_{i=1}^{n}\left(b^2_i+b_i\cdot a_i+a_i^2\right)}_{(I)}-\underbrace{\frac{1}{4\cdot n^2}\left[\sum\limits_{i=1}^{n}\left(a_i-b_i\right)\right]^2}_{(II)}.
\end{equation}
Considering that:
\begin{eqnarray*}
\mu _i^2 + \sigma _i^2 &=& {\left( {\frac{{{b_i} + {a_i}}}{2}} \right)^2} + \frac{{{{({b_i} - {a_i})}^2}}}{{12}} = \frac{{{{({b_i} + {a_i})}^2}}}{4} + \frac{{{{({b_i} - {a_i})}^2}}}{{12}} = \\
 &=& \frac{{3b_i^2 + 3a_i^2 + 6{b_i}{a_i} + b_i^2 + a_i^2 - 2{b_i}{a_i}}}{{12}} = \\
 &=& \frac{{4b_i^2 + 4a_i^2 + 4{b_i}{a_i}}}{{12}} = \frac{{b_i^2 + {b_i}\cdot{a_i} + a_i^2}}{3}
\end{eqnarray*}
the term (I) of eq. (\ref{var_Bi}) can be expressed as follows:
$$
\frac{1}{3\cdot n}\sum\limits_{i=1}^{n}\left(b^2_i+b_i\cdot a_i+a_i^2\right)=\frac{1}{n}\sum\limits_{i=1}^{n}\left(\mu_i^2+\sigma_i^2\right) .
$$
The term  $(II)$ is clearly $\mu^2$, indeed:
$$
\frac{1}{4\cdot n^2}\left[\sum\limits_{i=1}^{n}\left(a_i-b_i\right)\right]^2=\left[\frac{1}{n}\sum\limits_{i=1}^{n}\frac{\left(a_i-b_i\right)}{2}\right]^2=\left[\frac{1}{n}\sum\limits_{i=1}^{n}\mu_i\right]^2=\mu^2 .
$$
Thus, $S^2$ in eq. (\ref{var_Bi}) corresponds to eq. (\ref{s2_mix}), indeed:
\begin{equation}\label{var_MIX}
S^2=(I)-(II)=\frac{1}{n}\sum\limits_{i=1}^{n}\left(\mu_i^2+\sigma_i^2\right)-\mu^2=\sigma^2 .
\end{equation}
The same correspondences also hold for the mean and the variance of the other numerical modal symbolic variables.\\

\section{\citet{BerGoup2000} approach to basic statistics}
A bit of notation
\begin{description}
  \item[Interval data] $Y_1$ and $Y_2$ are interval-valued data. $Y_1(i)=[a_{i1},b_{i1}]$ and $Y_2(i)=[a_{i2},b_{i2}]$.
  \item[Histogram data] $Y_1$ and $Y_2$ are histogram-valued data.
  $$Y_1(i)=\{([a_{i1,1},b_{i1,1}],\pi_{i1,1}),\ldots,([a_{i1,h_{i1}},b_{i1,h_{i1}}],\pi_{i1,h_{i1}})\}$$  such that $$\sum\limits_{r=1}^{h_{i1}}\pi_{i1,h_{i1}}=1$$ and
  $$Y_2(i)=\{([a_{i2,1},b_{i2,1}],\pi_{i2,1}),\ldots,([a_{i2,h_{i2}},b_{i2,h_{i2}}],\pi_{i2,h_{i2}})\}$$  such that $$\sum\limits_{r=1}^{h_{i2}}\pi_{i2,h_{i2}}=1.$$
  $h_{i1}$ is not necessarily equal to $h_{i2}$.
\end{description}
\citet{BerGoup2000} and \citep{bildid03}
\begin{description}
  \item[Univariate statistics for intervals] \citet{BerGoup2000} consider each interval as a uniform distributions. Under this hypothesis the \textbf{mean} is \\
  $$\bar{Y}=\frac{1}{n}\sum\limits_{i = 1}^n {\frac{{{b_i} + {a_i}}}{2}} $$
  being $\mu_i=\frac{{b_i} + {a_i}}{2}$ then
  $$\bar{Y}=\frac{1}{n}\sum\limits_{i = 1}^n {\mu_i} $$
  the \textbf{variance} is
  $$S^2=\frac{1}{{3n}}\sum\limits_{i = 1}^n {\left( {b_i^2 + {b_i}{a_i} + a_i^2} \right)} - {\frac{1}{{4{n^2}}}{{\left[ {\sum\limits_{i = 1}^n {\left( {{b_i} + {a_i}} \right)} } \right]}^2}} $$
  being  $\sigma_i=\sqrt{\frac{({b_i} -{a_i})^2}{12}}$ and being
  $$\mu_i^2+\sigma_i^2=\frac{{{{\left( {{b_i} + {a_i}} \right)}^2}}}{4} + \frac{{{{\left( {{b_i} - {a_i}} \right)}^2}}}{{12}} = \frac{{4b_i^2 + 4a_i^2 + 4{a_i}{b_i}}}{{12}} = \frac{1}{3}\left( {b_i^2 + {a_i}{b_i} + a_i^2} \right)$$
  then
  $$S^2=\frac{1}{{n}}\sum\limits_{i = 1}^n{(\mu_i^2+\sigma_i^2)}-\bar{Y}^2.$$
  \item[Bivariate statistics for intervals] In the bivariate case, \citet{BerGoup2000}  assume that, if the individual is observed for two interval-valued variables, the joint distribution whose  marginals are the two uniforms is derived under an (implicit) independence assumption, i.e., given $Y_1(i)\sim\mathcal{U}(a_{i1},b_{i1})$ and $Y_2(i)\sim\mathcal{U}(a_{i2},b_{i2})$  the within covariance $COV(Y_1(i), Y_2(i))=0$ thus $(Y_1\cap Y_2)(i)=Y_1(i)\cdot Y_2(i)=\mathcal{U}(a_{i1},b_{i1})\cdot\mathcal{U}(a_{i2},b_{i2}).$\\
      Using the same approach of \citet{Bill2008IASC}, we can decompose the cross-variation into a Within and a between component as follows
      $$CSW=\sum\limits_{i=1}^n COV(Y_1(i), Y_2(i))$$ that is equal to $0$ because each  $COV(Y_1(i), Y_2(i))=0$, and a between that is equal to
      $$CSB=\sum\limits_{i=1}^n(\mu_{i1}-\bar{Y}_1)(\mu_{i2}-\bar{Y}_2)=\sum\limits_{i=1}^n\mu_{i1}\mu_{i2}-n\bar{Y}_1\bar{Y}_2$$
      thus
      \begin{equation}\label{betr_goup}
        CST=CSW+CSB=0+CSB=\sum\limits_{i=1}^n\mu_{i1}\mu_{i2}-n\bar{Y}_1\bar{Y}_2.
      \end{equation}
      Considering the independence and the uniform assumption the \textbf{covariance} of a set of bi-variate intervals is
      $$COV=\frac{CST}{n}=\frac{1}{n}\sum\limits_{i=1}^n\mu_{i1}\mu_{i2}-\bar{Y}_1\bar{Y}_2.$$

  \item[Univariate statistics for histograms] Is analogue to the intervals. The \textbf{mean}
  $$\bar{Y}=\frac{1}{{2n}}\sum\limits_{i = 1}^n { \underbrace{\sum\limits_{h = 1}^{{H_i}} {\left( {{b_{ih}} + {a_{ih}}} \right)}{\pi _{ih}} }_{2\cdot\mu_i}} = \frac{1}{{n}}\sum\limits_{i = 1}^n\mu_i$$
  while the  \textbf{variance} is
 $$S^2=\int\limits_{ - \infty }^{ + \infty } {{{\left( {y - \bar Y} \right)}^2}f(y)dy}  $$
 Using the approach suggested by \citet{Bill2008IASC} we can divide $nS$ into a within and a between part. The within sum of squares $SSW$ is the sum of internal variability:
 $$SSW=\sum\limits_{i=1}^n\sigma^2_i=\sum\limits_{i=1}^n{\int\limits_{ - \infty }^{ + \infty } {{{\left( {y - \mu_i} \right)}^2}f_i(y)dy}  }={\sum\limits_{i = 1}^n {\left[ {\int\limits_{ - \infty }^{ + \infty } {{y^2}{f_i}(y)dy}  - {{ {{\mu _i}} }^2}} \right]} }$$
 where considering that $f_i(y)=\frac{\pi_{ih}}{b_{ih}-a_{ih}}$ if $a_{ih}\leq y \leq b_{ih}$, we have
 $$
\sum\limits_{h = 1}^{{H_i}} {\frac{{{\pi _{ih}}}}{{{b_{ih}} - {a_{ih}}}}\int\limits_{{a_{hi}}}^{{b_{ih}}} {{y^2}dy} }  = \sum\limits_{h = 1}^{{H_i}} {\frac{{{\pi _{ih}}}}{{{b_{ih}} - {a_{ih}}}}} \frac{1}{3}\left[ {b_{hi}^3 - a_{hi}^3} \right]
 = \frac{1}{3}\sum\limits_{h = 1}^{{H_i}} {{\pi _{ih}}} \left[ {a_{ih}^2 + {b_{ih}}{a_{ih}} + b_{ih}^2} \right]
$$
thus
$$SSW=\sum\limits_{i=1}^n\sigma^2_i=\frac{1}{3}\sum\limits_{i=1}^n\sum\limits_{h = 1}^{{H_i}} {{\pi _{ih}}} \left[ {a_{ih}^2 + {b_{ih}}{a_{ih}} + b_{ih}^2} \right]-\sum\limits_{i=1}^n\mu_i^2.$$
the between sum of squares is
$$SSB=\sum\limits_{i=1}^n\left(\mu_i-\bar{Y}\right)^2=\sum\limits_{i=1}^n\mu_i^2-n\bar{Y}^2$$
thus
$$SST=SSW+SSB=\frac{1}{3}\sum\limits_{i=1}^n\sum\limits_{h = 1}^{{H_i}} {{\pi _{ih}}} \left[ {a_{ih}^2 + {b_{ih}}{a_{ih}} + b_{ih}^2} \right]-\sum\limits_{i=1}^n\mu_i^2+\sum\limits_{i=1}^n\mu_i^2-n\bar{Y}^2$$
thus the variance is
$$S^2=\frac{SST}{n}= \frac{1}{3n}\sum\limits_{i=1}^n\sum\limits_{h = 1}^{{H_i}} {{\pi _{ih}}} \left[ {a_{ih}^2 + {b_{ih}}{a_{ih}} + b_{ih}^2} \right]-\bar{Y}^2$$
  \item[Bivariate statistics for histograms] Also in this case, \citet{BerGoup2000} assume the internal independence for bivariate histogram-valued description, i.e., they assume that that  it is assumed that $COV(Y_1(i), Y_2(i))=0$. Therefore, using the \citet{Bill2008IASC} approach we can write:
      $$CSW=\sum\limits_{i=1}^n COV(Y_1(i), Y_2(i))$$ that is equal to $0$ because each  $COV(Y_1(i), Y_2(i))$, and a between that is equal to
      $$CSB=\sum\limits_{i=1}^n(\mu_{i1}-\bar{Y}_1)(\mu_{i2}-\bar{Y}_2)=\sum\limits_{i=1}^n\mu_{i1}\mu_{i2}-n\bar{Y}_1\bar{Y}_2$$
      thus
      $$CST=CSW+CSB=0+CSB=\sum\limits_{i=1}^n\mu_{i1}\mu_{i2}-n\bar{Y}_1\bar{Y}_2.$$
\end{description}
A general drawback is that given two variables $Y_1$ and $Y_2$ such that $Y_1(i)=Y_2(i)$ for each $i=1,\ldots,n$, looking at the formulas it easy to observe that
even if $S^2(Y_1)=S^2(Y_2)$,
\begin{equation}\label{problem1}
CST(Y_1,Y_2)\neq n\cdot S^2(Y_1)\;\;or\;\;CST(Y_1,Y_2)\neq n\cdot S^2(Y_2).
\end{equation}

\section{\citet{BilDid2006} 2006 approach to basic statistics}
\begin{description}
  \item[Univariate statistics for intervals] They are the same of \citet{BerGoup2000} under the hypothesi of uniform distribution in each interval-valued description.
  \item[Bivariate statistics for intervals] Are a particular modification of the \citet{BerGoup2000}, but without solving some drawbacks (the covariance of two identical interval variables is different from the variance of the two variables.
  \item[Univariate statistics for histograms] They are the same of \citet{BerGoup2000} under the hypothesi of uniform distribution in each histogram-valued description.
  \item[Bivariate statistics for histograms] Are a particular modification of the \citet{BerGoup2000}, but without solving some drawbacks (the covariance of two identical interval variables is different from the variance of the two variables. Another problem arise, if the histograms are the same but have a different bin partition.
      
\end{description}

\section{\citet{Bill2008IASC} 2008 approach to basic statistics}
\paragraph[Billard 2008]define interval descriptions and histogram descriptions
\begin{description}
  \item[Univariate statistics for intervals]Are the same of \citet{BerGoup2000} under the hypothesi of uniform distribution in each interval-valued description.
  \item[Bivariate statistics for intervals] The main novelty of the \citet{Bill2008IASC} approach is related to the definition of the cross-variation between two interval-valued variables. Indeed, it is assumed that
      \begin{equation}\label{AssumptionBill}
       CSW=\sum\limits_{i=1}^n COV(Y_1(i), Y_2(i))=\sum\limits_{i=1}^n(b_{i1}-a_{i1})(b_{i2}-a_{i2})/12.
      \end{equation}
       The assumption that $COV(Y_1(i), Y_2(i))=(b_{i1}-a_{i1})(b_{i2}-a_{i2})/12$ is related to an  assumption (not declared in the paper) of perfect positive internal correlation between of the two uniforms describing the $i$-th individual. In fact, being $\sigma_{i1}^2=\frac{(b_{i1}-a_{i1})^2}{12}$ and $\sigma_{i2}^2=\frac{(b_{i2}-a_{i2})^2}{12}$ and denoting with $CORR(Y_1(i), Y_2(i))$ the correlation between $Y_1(i)$ and $Y_2(i)$, we can write:
       $$CORR(Y_1(i), Y_2(i))=\frac{COV(Y_1(i), Y_2(i))}{\sigma_{i1}\sigma{i2}}=1\rightarrow$$ $$\rightarrow COV(Y_1(i), Y_2(i))=\sigma_{i1}\sigma_{i2}=\frac{(b_{i1}-a_{i1})(b_{i2}-a_{i2})}{12}$$
       Naturally the between component remain the same, i.e.
       $$CSB=\sum\limits_{i=1}^n(\mu_{i1}-\bar{Y}_1)(\mu_{i2}-\bar{Y}_2)=\sum\limits_{i=1}^n\mu_{i1}\mu_{i2}-n\bar{Y}_1\bar{Y}_2$$
      thus
      $$CST=CSW+CSB=\sum\limits_{i=1}^n(b_{i1}-a_{i1})(b_{i2}-a_{i2})/12+\sum\limits_{i=1}^n\mu_{i1}\mu_{i2}-n\bar{Y}_1\bar{Y}_2$$
       and being
       \[\begin{array}{l}
  \sum\limits_{i = 1}^n {\frac{{({b_{i1}} - {a_{i1}})({b_{i2}} - {a_{i2}})}}{{12}}}  + \sum\limits_{i = 1}^n {\frac{{({b_{i1}} + {a_{i1}})({b_{i2}} + {a_{i2}})}}{4}}  - n{{\bar Y}_1}{{\bar Y}_2} = \\
 = \sum\limits_{i = 1}^n {\left[ {\frac{{({b_{i1}} - {a_{i1}})({b_{i2}} - {a_{i2}})}}{{12}} + \frac{{({b_{i1}} + {a_{i1}})({b_{i2}} + {a_{i2}})}}{4} - {{\bar Y}_1}{{\bar Y}_2}} \right]}  = \\
 = \sum\limits_{i = 1}^n {\left[ {\frac{1}{6}\left( {2{b_{i1}}{b_{i2}} + {b_{i1}}{a_{i2}} + {a_{i1}}{b_{i2}} + 2{a_{i1}}{a_{i2}}} \right) - {{\bar Y}_1}{{\bar Y}_2}} \right]}
\end{array}\]
we have that
$$CST=CSW+CSB=\sum\limits_{i = 1}^n {\left[ {\frac{1}{6}\left( {2{b_{i1}}{b_{i2}} + {b_{i1}}{a_{i2}} + {a_{i1}}{b_{i2}} + 2{a_{i1}}{a_{i2}}} \right) + {{\bar Y}_1}{{\bar Y}_2}} \right]}$$
that, being independent for translations, it can be expressed as in \cite{Bill2008IASC} as follows
\[\begin{array}{l}
CST = CSW + CSB = \\
 = \frac{1}{6}\sum\limits_{i = 1}^n {\left[ \begin{array}{l}
2({b_{i1}} - {{\bar Y}_1})({b_{i2}} - {{\bar Y}_2}) + ({b_{i1}} - {{\bar Y}_1})({a_{i2}} - {{\bar Y}_2}) + \\
 + ({a_{i1}} - {{\bar Y}_1})({b_{i2}} - {{\bar Y}_2}) + 2({a_{i1}} - {{\bar Y}_1})({a_{i2}} - {{\bar Y}_2})
\end{array} \right]}
\end{array}\]
        Even if it is not discussed in the paper \cite{Bill2008IASC}, this formulation seems a numerical trick for solving the problem in Eq. (\ref{problem1}).
       In fact in this case it is easy to show that, given two interval-valued variables $Y_1$ and $Y_2$ such that $Y_1(i)=Y_2(i)$ for each $i=1,\ldots,n$,
       $$CST(Y_1,Y_2)= n\cdot S^2(Y_1)=n\cdot S^2(Y_2).$$
       {\bf However, this is true only for interval valued data} (as we see in a while).
  \item[Univariate statistics for histograms] Are the same of \citet{BerGoup2000} under the hypothesis of uniform distribution in each bin of the histogram-valued description.
  \item[Bivariate statistics for histograms] In this case \citet{Bill2008IASC}, without any justification, extend the covariance using a weighted formulation of equation that implies, for each couple of bins of the two histograms that describe the $i$-th individual, as follows:
      \[\begin{array}{l}
COV({Y_1},{Y_2}) = \\\frac{1}{{6n}}\sum\limits_{i = 1}^n {\sum\limits_{r = 1}^{{h_{i1}}} {\sum\limits_{s = 1}^{{h_{i2}}} {\left\{ {\left[ {2\left( {{b_{i1,r}} - {{\bar Y}_1}} \right)\left( {{b_{i2,s}} - {{\bar Y}_2}} \right) + \left( {{b_{i1,r}} - {{\bar Y}_1}} \right)\left( {{a_{i2,s}} - {{\bar Y}_2}} \right) + } \right.} \right.} } } \\
\left. {\left. { + \left( {{a_{i1,r}} - {{\bar Y}_1}} \right)\left( {{b_{i2,s}} - {{\bar Y}_2}} \right) + 2\left( {{a_{i1,r}} - {{\bar Y}_1}} \right)\left( {{a_{i2,s}} - {{\bar Y}_2}} \right)} \right]{\pi _{i1,r}}{\pi _{i2,s}}} \right\}
\end{array}\]
However, this formulation does not solve the problem in Eq. (\ref{problem1}), and further it is sensible to a recodify of the histograms that does not change the density. We see this with two examples. Further, when the the number of bins increases $h_{..}\rightarrow+\infty$, in general $\pi_{...}\rightarrow0$, the consequence is that $COV\rightarrow  \frac{1}{n}\sum\limits_{i=1}^n\mu_{i1}\mu_{i2}-\bar{Y}_1\bar{Y}_2$. The proof is intuitive (because the bivariate histogram tends to coincide with the density of a bivariate distribution under independence assumption).
\paragraph{Example: problem in Eq. (\ref{problem1}) persists} We have the following dataset of 2 individuals described by 2 histogram variables.
\begin{table}[h]
  \centering
  \begin{tabular}{|c|c|c|}
     \hline
     i & $Y_1$ & $Y_2$ \\
     \hline
     1 & $\left\{([10,20],0.4),([20,30],0.6)\right\}$ & $\left\{([10,20],0.4),([20,30],0.6)\right\}$  \\
     2 & $\left\{([50,60],0.2),([60,70],0.8)\right\}$  & $\left\{([50,60],0.2),([60,70],0.8)\right\}$  \\
     \hline
   \end{tabular}
    \begin{tabular}{|p{13cm}|}
     \hline
     basic statistics \\
     \hline
    \[\begin{array}{l}
\overline {{Y_1}}  = \frac{1}{2}\left( {\frac{{20 + 10}}{2}0.4 + \frac{{30 + 20}}{2}0.6 + \frac{{60 + 50}}{2}0.8 + \frac{{70 + 60}}{2}0.8} \right) = 42\\
\overline {{Y_2}}  = \frac{1}{2}\left( {\frac{{20 + 10}}{2}0.4 + \frac{{30 + 20}}{2}0.6 + \frac{{60 + 50}}{2}0.8 + \frac{{70 + 60}}{2}0.8} \right) = 42\\
S_1^2 = \frac{1}{{3 \cdot 2}}\left\{ \begin{array}{l}
\left[ {{{10}^2} + 10 \cdot 20 + {{20}^2}} \right]0.4 + \left[ {{{20}^2} + 20 \cdot 30 + {{30}^2}} \right]0.6 + \\
\left[ {{{50}^2} + 50 \cdot 60 + {{60}^2}} \right]0.2 + \left[ {{{60}^2} + 60 \cdot 70 + {{70}^2}} \right]0.8
\end{array} \right\} - {42^2} = 469,\bar 3\\
S_2^2 = \frac{1}{{3 \cdot 2}}\left\{ \begin{array}{l}
\left[ {{{10}^2} + 10 \cdot 20 + {{20}^2}} \right]0.4 + \left[ {{{20}^2} + 20 \cdot 30 + {{30}^2}} \right]0.6 + \\
\left[ {{{50}^2} + 50 \cdot 60 + {{60}^2}} \right]0.2 + \left[ {{{60}^2} + 60 \cdot 70 + {{70}^2}} \right]0.8
\end{array} \right\} - {42^2} = 469,\bar 3\\
\\
COV({Y_1},{Y_2}) = \\
 = \frac{1}{{6 \cdot 2}}\left\{ \begin{array}{l}
\left[ \begin{array}{l}
2(10 - 42)(10 - 42) + (20 - 42) \cdot (10 - 42) + \\
 + (10 - 42) \cdot (20 - 42) + 2(20 - 42) \cdot (20 - 42)
\end{array} \right]0.4 \cdot 0.4 + \\
 + \left[ \begin{array}{l}
2(10 - 42)(20 - 42) + (10 - 42) \cdot (30 - 42) + \\
 + (20 - 42) \cdot (20 - 42) + 2(20 - 42) \cdot (30 - 42)
\end{array} \right]0.4 \cdot 0.6 + \\
....
\end{array} \right\}=449.\bar{3}
\end{array}\]\\
     \hline
   \end{tabular}
   \caption{EX1}\label{example1}
\end{table}

\paragraph{Example: COV is not invariant to recodify of the same histogram into different bins} If we take $Y_2$ and rewrite it by splitting the bins into two parts we do not change the density or the cumulative function associated with each histogram. However, $COV(Y_1,Y_2)$ changes. In order to show this, in Tab. \ref{example2}, we have recodified the second variable, as follows:\\
$$\begin{array}{l}Y_2(1)=\left\{([10,20],0.4),([20,30],0.6)\right\}=\\=\left\{([10,15],0.2),([15,20],0.2),([20,25],0.3),([25,30],0.3)\right\}\end{array}$$
and
$$\begin{array}{l}Y_2(2)=\left\{([50,60],0.2),([60,70],0.8)\right\}=\\=\left\{([50,55],0.1),([55,60],0.1),([60,65],0.4),([65,70],0.4)\right\}\end{array}.$$
Thus, rewriting the histogram without changing its distribution or its density, $COV$ changes, but not the univariate statistics. Continuing to bisecting bins, it is easy to show that $COV({Y_1},{Y_2})$ tend to the covaraince of the biveriate distribution of the means of the histograms that is equal to 441. In appendix, the Matlab code of the numerical proof.
\begin{table}[h]
  \centering
  \begin{tabular}{|c|c|c|}
     \hline
     i & $Y_1$ & $Y_2$ \\
     \hline
     1 & $\left\{([10,20],0.4),([20,30],0.6)\right\}$ & $\left\{([10,15],0.2),[15,20],0.2),([20,25],0.3)[25,30],0.3)\right\}$  \\
     2 & $\left\{([50,60],0.2),([60,70],0.8)\right\}$  & $\left\{([50,55],0.1),([55,60],0.1),([60,65],0.4),([65,70],0.4)\right\}$  \\
     \hline
   \end{tabular}
    \begin{tabular}{|p{13cm}|}
     \hline
     basic statistics \\
     \hline
    \[\begin{array}{l}
\overline {{Y_1}}  = 42\\
\overline {{Y_2}}  =  42\\
S_1^2 = 469,\bar 3\\
S_2^2 =  469,\bar 3\\
COV({Y_1},{Y_2}) =445.1\bar{6}
\end{array}\]\\
     \hline
   \end{tabular}
   \caption{EX2, the second variable is split but the histogram have always the same density function}\label{example2}
\end{table}

\end{description}

\newtheorem{thm}{Proposition}

\section*{Appendix} Code for proof related to example 1 and 2.\\
The function $cov_stat_billard(HM)$ computes the basic statistics for two histogram variables. HM is a structure where each element is a histogram.
\begin{lstlisting}
function res=cov_stat_billard(HM)
%this function compute the basic statistics of Billard 2008 IASC
n=size(HM,1);
ALL1=[];
ALL2=[];
meds=[];
for i=1:size(HM,1);

    ALL1=[ALL1; HM(i,1).h];
    ALL2=[ALL2; HM(i,2).h];
    meds(i,1)=sum(( HM(i,1).h(:,2)+HM(i,1).h(:,1))/2.*HM(i,1).h(:,3));
    meds(i,2)=sum(( HM(i,2).h(:,2)+HM(i,2).h(:,1))/2.*HM(i,2).h(:,3));
end
m1=1/n*sum((ALL1(:,2)+ALL1(:,1))/2.*ALL1(:,3));
m2=1/n*sum((ALL2(:,2)+ALL2(:,1))/2.*ALL2(:,3));
res.m1=m1;
res.m2=m2;

tmp1=0;
for i=1:n
    for r=1:size(HM(i,1).h,1)
        tmp=(HM(i,1).h(r,1)-m1)^2+...
            (HM(i,1).h(r,1)-m1)*(HM(i,1).h(r,2)-m1)+...
            (HM(i,1).h(r,2)-m1)^2;
        tmp1=tmp1+1/(3*n)*tmp*HM(i,1).h(r,3);

    end
end
var1=tmp1;
res.var1=var1;
tmp2=0;
for i=1:n
    for r=1:size(HM(i,2).h,1)
        tmp=(HM(i,2).h(r,1)-m2)^2+...
            (HM(i,2).h(r,1)-m2)*(HM(i,2).h(r,2)-m1)+...
            (HM(i,2).h(r,2)-m2)^2;
        tmp2=tmp2+1/(3*n)*tmp*HM(i,2).h(r,3);

    end
end
var2=tmp2;
res.var2=var2;

covar=0;
tmp=0;
for i=1:n

    for r=1:size(HM(i,1).h,1)
        for t=1:size(HM(i,2).h,1)
            tmp2=2*(HM(i,1).h(r,1)-m1)*(HM(i,2).h(t,1)-m2)+...
                (HM(i,1).h(r,1)-m1)*(HM(i,2).h(t,2)-m2)+...
                (HM(i,1).h(r,2)-m1)*(HM(i,2).h(t,1)-m2)+...
                2*(HM(i,1).h(r,2)-m1)*(HM(i,2).h(t,2)-m2);

            tmp=tmp+1/(6*n)*tmp2*HM(i,1).h(r,3)*HM(i,2).h(t,3);
        end
    end
end
covar=tmp;
res.covar=covar;
res.covCE=meds(:,1)'*meds(:,2)*1/n-mean(meds(:,1))*mean(meds(:,2));
\end{lstlisting}
The next script starts from a configuration of histograms, thus split each bin in two equal parts and recompute basic statistics.
\begin{lstlisting}
% define 2x2 histogram-valued table 
H11=[10 20 0.4;...
    20 30 0.6];
H21=[50 60 0.2;...
    60 70 0.8];
H12=[10 20 0.4;...
    20 30 0.6];
H22=[50 60 0.2;...
    60 70 0.8];

% set up the structure HM
HM(1,1).h=H11;
HM(1,2).h=H12;
HM(2,1).h=H21;
HM(2,2).h=H22;

res=cov_stat_billard(HM);
res.covar
%now we do 5 splits and recompute basic statistics
for i=1:5
    for r=1:size(HM,1)
        for s=1:size(HM,2)
            tmp=HM(r,s).h;
            %%
            tmp2=[];
            for k=1:size(HM(r,s).h,1);
                uno=HM(r,s).h(k,1);
                tre=HM(r,s).h(k,2);
                due=(tre+uno)/2;
                pp=HM(r,s).h(k,3)/2;
                tmp2=[tmp2;[uno due pp; due tre pp]];

            end
            HM(r,s).h=tmp2;
        end
    end
    res=cov_stat_billard(HM);
    res.covar
    res.covCE
end
\end{lstlisting}

\bibliographystyle{plain}

\end{document}